\documentclass[12pt]{iopart}
\usepackage{graphicx}
\usepackage{color} 
\begin{document}
\title{High reflectivity grating waveguide coatings for 1064\,nm}
\author{A~Bunkowski, O~Burmeister, D~Friedrich, K~Danzmann and R~Schnabel}
\address{Max-Planck-Institut f\"ur Gravitationsphysik
 (Albert-Einstein-Institut) and Institut f\"ur Gravitationsphysik, Leibniz Universit\"at Hannover,
 Callinstr. 38, 30167 Hannover, Germany}
\ead{alexander.bunkowski@aei.mpg.de}
\begin{abstract}
We propose thin single-layer grating waveguide structures to be used as
high-reflectivity, but low thermal noise, alternative to conventional coatings
for gravitational wave detector test mass mirrors. Grating waveguide (GWG)
coatings can show a reflectivity of up to 100\% with  an overall thickness of
less than a wavelength. We theoretically investigate GWG coatings for 1064\,nm
based on tantala (Ta$_2$O$_5$) on a Silica substrate focussing on broad
spectral response and low thickness.
\end{abstract}
\maketitle
\section{Introduction}
Dedicated research during the last few years has revealed that thermally driven
motion of the test masses, so-called thermal noise~\cite{thermal}
is larger than foreseen in future gravitational wave detectors.
A major, but previously underestimated, contribution is given by the multilayer
dielectric coating stacks of the high reflectivity test mass
mirrors~\cite{coatingnoise,coatingnoise2}.
This currently limits the design sensitivity of the Advanced LIGO
detector~\cite{advligo}.
In conventional schemes, up to 40 layers of Ta$_2$O$_5$ and SiO$_2$ with an
optical thickness of a quarter wavelength are needed to reach high
reflectivities sufficiently close to 100\%.
The thermal noise of the coating is due to the mechanical loss  of the layers
with a dominant contribution from Ta$_2$O$_5$.
New concepts are required that have less loss but still achieve the required
high reflectivity.
One approach being pursued is to design an alternative multilayer system
deviating from the classical quarter wave design and containing less
Ta$_2$O$_5$ \cite{agresti}.
Doping of Ta$_2$O$_5$ with TiO$_2$ has also been investigated and a reduction
of the loss by a factor of 1.5 was observed~\cite{doping}.
%
Another approach is to avoid high reflection coatings at all by the use of
corner reflectors which employ total internal reflection instead of multiple
interference at different layers to reach high reflectivity~\cite{corner}.
However, in this case thermo-refractive noise which results from a temperature
dependent refractive index and also thermal lensing are increased due to the
large optical path length in the substrate material.

Grating waveguide structures~\cite{gws} provide another possibility to
construct high reflectivity devices.
The interest of earlier work on grating waveguides lay mainly in  narrow-band
(highly resonant) devices for applications in optical
filtering~\cite{narrowfilter} and optical switching~\cite{optswitch}. However,
grating waveguide structures can also provide  broad-band (weakly resonant)
reflectors.
This turns them into  interesting candidates for test mass coatings in
gravitational wave detectors, because only a very small amount dielectric
coating material is required with a corresponding considerable reduction in
coating thermal noise.

\section{Resonant grating waveguide structures}
\begin{figure}[htb]
    \centerline{\includegraphics[width=11cm]{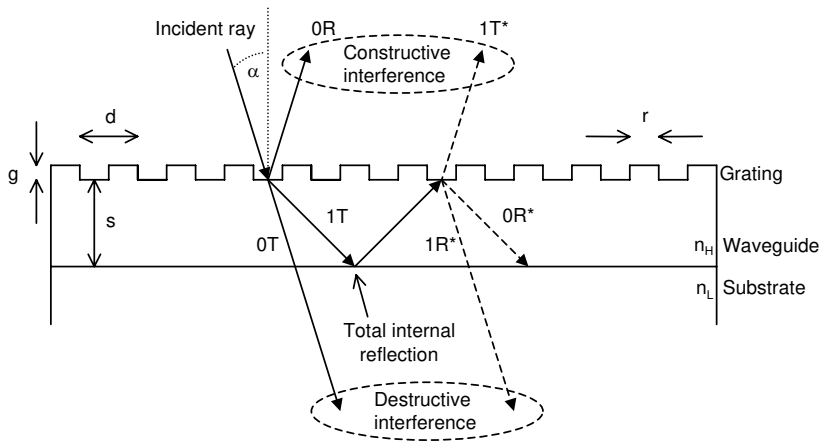}}
    \caption{Schematic of a grating waveguide structure in a simplified ray
    picture. For clarity a non zero angle of incidence and only one first order
    transmission is shown.}
    \label{fig:gws}
\end{figure}
The remarkable property of a grating waveguide (GWG) is that it can show a
reflectivity of 100\,\% for a given optical wavelength $\lambda$ despite its
thickness of typically less than a wavelength. For an extensive overview of
grating waveguides we refer to Ref.~\cite{gwsreview}.
In the simplest case a GWG consists of a substrate material with low refractive
index $n_{\mathrm{L}}$ followed by a waveguide layer with high refractive index
$n_{\mathrm{H}}$ which has periodic corrugation with period~$d$ as shown in
figure~\ref{fig:gws}.
A simplified ray picture~\cite{gws} can be used to understand its behavior.
The structure can be designed such that light incident onto the grating will
only produce one diffraction order in reflection (0R) but three diffraction
orders in transmission (0T and $\pm$1T).
(For clarity the -1T order has been omitted in figure~\ref{fig:gws}.)

The first order beams are coupled into a layer of high refractive material
where they are stored due to total internal reflection.
Light inside the waveguide is also coupled out via the grating.
For a proper choice of grating parameters and incident angle the light coupled
out from the layer (1R*) interferes destructively with the zero order
transmitted beam (0T) and the device is a perfect reflector.

The possible parameter range for the period $d$ depends on the angle of
incidence  $\alpha$, the (vacuum) laser wavelength $\lambda$ ($\lambda_0$), and
the refractive indices $n_{\mathrm{L}}$ and $n_{\mathrm{H}}$ and can be
calculated from the grating equation
\begin{equation}\label{eq:gr}
    \sin \alpha + \sin \beta_m = m\lambda/d,
\end{equation}
where $\beta_m$ is the angle of the $m$th diffraction order. For test mass
mirrors in gravitational wave detector Michelson interferometers the angle of
incidence is typically restricted to $\alpha=0$.
To ensure that only the  $m=0$ order is allowed in reflection,
\begin{equation} \label{eq:cond1}
d<\lambda_0
\end{equation}
has to hold.
Another condition is that only $|m|\leq1$ orders in transmission
exist from which follows that
\begin{equation}\label{eq:cond2}
    1 < d \frac{n_{\mathrm{H}}}{\lambda_0}<2.
\end{equation}
Total internal reflection of the first order at the boundary of
the waveguide and the substrate material is ensured if
\begin{equation}\label{eq:cond3}
    d<\lambda_0/n_{\mathrm{L}}.
\end{equation}

A resonant grating waveguide structure has analogous behavior to a Fabry-Perot
resonator: with decreasing coupling to the waveguide the finesse of the
structure increases~\cite{gws}.
For high reflectors in GW detectors high finesse structures are
disadvantageous, because small deviations from the design parameter would
dramatically decrease the reflectivity for the desired wavelength $\lambda_0$.
Additionally the power built up inside a high finesse waveguide
could be a problem for high-power laser interferometers.
Accordingly, a broadband resonance is desired for the high
reflector.

\section{Spectral response of waveguide coatings}
Using rigorous coupled wave (RCW) analysis~\cite{rcw} it is possible to
calculate the optical properties of the structure.
Design considerations for binary gratings must include groove depth $g$,
waveguide thickness $s$ and ridge width $r$, see figure~\ref{fig:gws}, in
addition to the before mentioned period $d$ and refractive indices
$n_{\mathrm{L}}$ and $n_{\mathrm{H}}$.
The goal is to design a
broad band reflection being less sensitive to fabrication tolerances and
avoiding the problem of strong light power buildup in the waveguide.
Here we restrict ourselves to $n_{\mathrm{H}}=2.04$ and $n_{\mathrm{L}}=1.45.$
This corresponds to tantala 
and fused silica which are the favorite high index coating material and  test
mass material respectively~\cite{coatingnoise}.

\begin{figure}[htb]
    \centerline{\includegraphics[width=16cm]{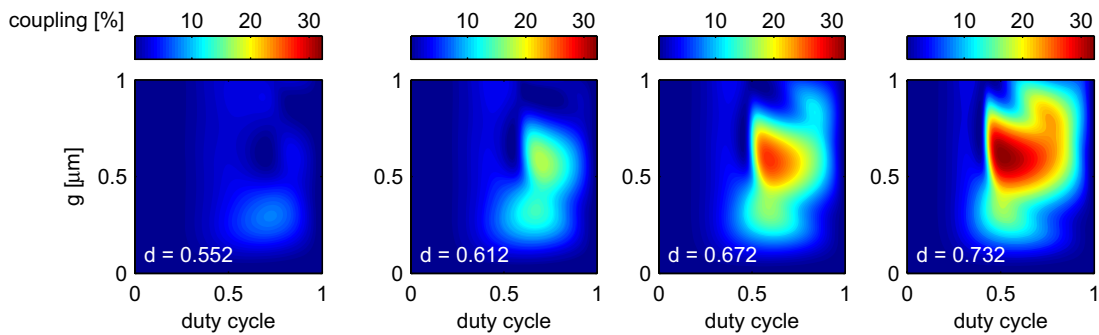}}
    \caption{Coupling to the waveguide for the $\pm1$T rays (color-coded) versus groove depth $g$ and
    fill factor $r/d$ for TM illumination and selected values of $d$.}
    \label{fig:transTM}
\end{figure}

\begin{figure}[htb]
    \centerline{\includegraphics[width=8cm]{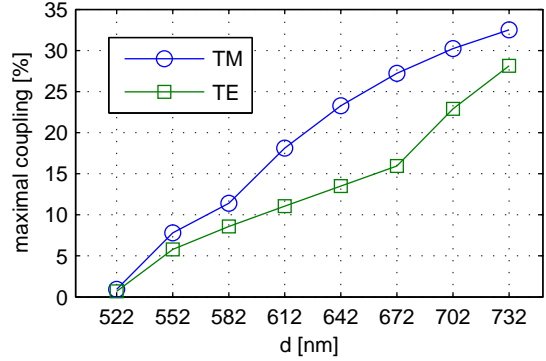}}
    \caption{Maximum achievable coupling for TM and TE polarization versus grating
    period $d$. For each point the groove depth was varied between 0 and 1$\mu$m
    and the fill factor between 0 and 1.}
    \label{fig:bestd}
\end{figure}

According to (\ref{eq:cond1}) -- (\ref{eq:cond3}) the following constraints
apply to the period when one assumes the commonly uses Nd:YAG laser wavelength
of $\lambda_0=1064$\,nm:
\begin{equation}\label{eq:allcond}
    521\,\mathrm{nm} < d < 734\,\mathrm{nm}.
\end{equation}
For a broadband response, the coupling to the waveguide which corresponds to
the diffraction efficiency of the $\pm$1T ray should be maximized.
It only depends on the grating properties $g$ and $r/d$ but not on the
thickness $s$ of the waveguiding layer.
Figure~\ref{fig:transTM} shows how the coupling depends on the groove depth $g$
and fill factor $(r/d)$ for selected values of $d$ for TM (magnetic field
vector is parallel to the grooves) illumination.
The plots indicate that the maximum coupling increases with increasing period
$d$.
This is illustrated in figure~\ref{fig:bestd} where we plotted the maximum
values of the coupling obtained when $g$ and $r/d$ were varied according to
figure~\ref{fig:transTM} versus grating period $d$ for TM and TE polarization.
Hence for the purpose of a broad-band reflection peak large values for the
grating period are favorable.
\begin{figure}[htb]
    \centerline{\includegraphics[width=14cm]{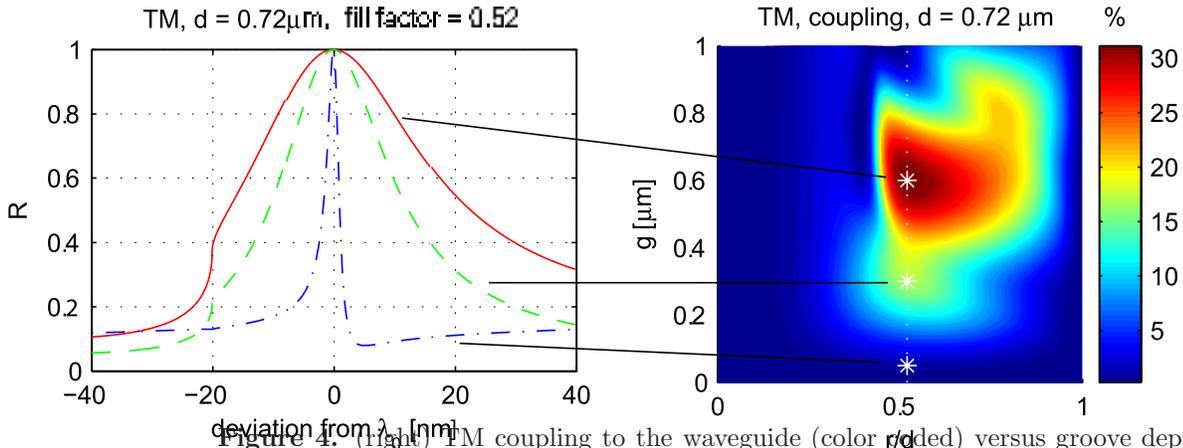}}
    \caption{(right) TM coupling to the waveguide (color coded) versus
    groove depth and fill factor. (left) Spectral response of three waveguide
    structures corresponding to the marked values in the right figure.}
    \label{fig:combi}
\end{figure}

The direct connection between coupling and spectral width of the resonance is
illustrated in figure~\ref{fig:combi}.
The right-hand side of the figure shows again the coupling to the waveguide
versus groove depth and fill factor for a specific grating period.
For three selected values of the groove depth (marked with three asterisks) and
fixed fill factor, we determined the optimal waveguide thickness $s_0$ for a
resonance peak around $\lambda_0=1064\,$nm.
On the left hand side of figure~\ref{fig:combi} we show the reflectivity versus
the deviation from $\lambda_0$ for the corresponding waveguide coating.


We note that for materials with higher refractive index than
$n_{\mathrm{H}}=2.04$ higher diffraction efficiencies (couplings to the
waveguide) and therefore even broader reflection peaks are
possible~\cite{Mateus1,Mateus2}.

\section{Thickness of the coating}
The crucial factor for coating thermal noise in gravitational wave detector
test masses is the overall thickness of the high index coating material.
To reach a reflectivity of $1-R = 10$\,ppm with a $\lambda/4$ stack of SiO$_2$
and Ta$_2$O$_5$, typically 40 layers are needed, adding up to
$20\lambda/(4n_{\mathrm{H}})\approx 2.6\,\mu$m overall tantala thickness.
In contrast to this, a grating waveguide mirror can get along with a tantala
thickness of much less than a wavelength.

\begin{figure}[htb]
    \centerline{\includegraphics[width=9cm]{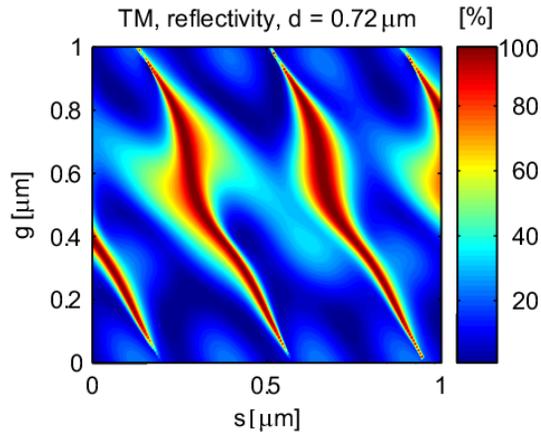}}
    \caption{Color coded TM-reflectivity of a waveguide structure versus groove
    depth and waveguide thickness $s$. The values for $d, r/d$ and $g$
    correspond to the dotted line in figure~\ref{fig:combi} (right).
    In this case 100\,\% reflectivity can be obtained for vanishing layer
    thickness at a groove depth of $g\approx 0.39\,\mu$m.}
    \label{fig:r}
\end{figure}

In addition, if the total thickness of tantala in the waveguide structure is to
be compared with a conventional mirror, one has to take into account that the
grating region is not uniformly filled.
Hence to first order approximation one can assume that
the coating thermal noise should be proportional to an effective tantala
thickness of $s+g(r/d)$.
The layer thickness $s$ determines the phase of the light travelling in the
waveguide and hence the resonance condition of the device.
The thickness $s_0$ for which a resonance occurs varies if the grating
parameters $g, r$ and $d$ are changed as illustrated exemplary in
figure~\ref{fig:r}, where the power reflectivity is plotted versus the groove
depth and the waveguide thickness.
One can see a periodic behavior of the reflectivity as $s$ varies as expected.
%
More interesting to note is that for a certain value of $g$ the 100\,\%
reflectivity resonance occurs at $s=0$.
Accordingly the grating itself can provide perfect reflection and no waveguide
layer is needed.
This is extremely useful since the amount of the high index material can be
greatly reduced.

An optimal design of a grating waveguide coating for gravitational wave
detectors will be a tradeoff between the broadest spectral response and the
smallest effective tantala thickness.
As an example we consider the GWS corresponding to the dashed (green) curve on
the left hand side of figure~\ref{fig:combi} which still has
$\Delta\lambda_{\mathrm{FWHM}}\approx 22\,$nm.
With $g=0.3\,\mu$m and $s\approx0.06\,\mu$m the effective tantala thickness is
only about $0.24\,\mu$m.
This suggests a  thermal noise reduction by more than an order of magnitude
compared to a conventional coating.

\section{Parameter tolerances}

\begin{figure}[htb]
    \centerline{\includegraphics[width=8cm]{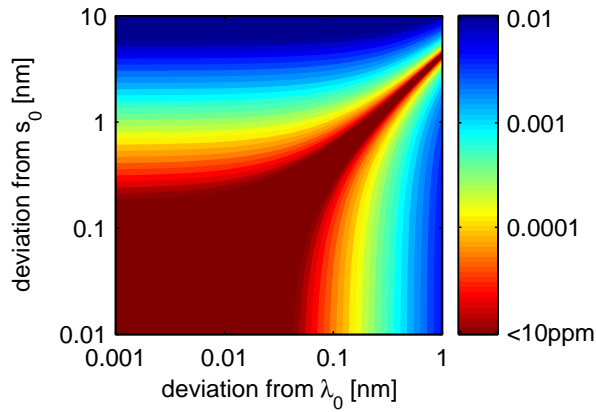}}
    \caption{Reflectivity plotted as 1-$R$ as function of deviation from optimal
    wavelength $\lambda_0$ and deviation from optimal waveguide thickness
    $s_0\approx239\,$nm. Other parameters for the grating: $g=0.6\,\mu$m;
    $r/d=0.52.$ }
    \label{fig:surface}
\end{figure}

When designing diffractive structures one also has to consider how accurately
grating parameters and layer thicknesses can be manufactured by state of the
art procedures and how strongly deviations from design values affect the
performance of the waveguide coating.
Here we consider the fabrication errors in the waveguide thickness and how they
can be compensated by tuning the laser wavelength.
figure~\ref{fig:surface} shows how the power reflectivity $R$ of a waveguide is
affected when the thickness of the waveguide or the wavelength of the laser
deviate from their optimal values $s_0$ and $\lambda_0$ respectively.
A typical power reflectivity requirement for GW detectors is $(1-R)<10\,$ppm.
Typical production accuracies of thin films are on the order of 1nm.
The deviation from $s_0$ could be compensated by tuning the laser wavelength a
small fraction of a nanometer.
Deviations in other grating parameters affect the reflectivity by a similar
way.

\section{Conclusion}
We have proposed a high reflectivity grating waveguide coating for advanced
gravitational wave detectors which can provide perfect reflection despite the
small amount of coating material that is needed.
This has great potential to lower the coating thermal noise of high
reflectivity mirrors.
Focussing on a laser wavelength of $\lambda_0=1064\,$nm and tantala as the coating
material we presented sample calculations of the spectral response of the
coating as well as the overall tantala thickness of the coating.
Our analysis was based on RCW analysis and assumed plane wave inputs as well as
infinite gratings.
Future theoretical work will include gaussian input beams and finite grating size effects.
On top of that more sophisticated designs of grating waveguide structures like
double periodic structures~\cite{Lemarchand} or double gratings~\cite{kappel}
will also be investigated.
Future experimental work will aim for a detailled characterization of such
devices as an alternative to conventional high reflectivity multilayer
dielectric coating stacks. An important issue will be the reduction of 
optical losses that may arise from writing errors during grating fabrication.

\ack This work has been supported by Deutsche Forschungsgemeinschaft within the
Sonderforschungsbereich TR7. The authors would also like to thank
T.~Clausnitzer, E.-B.~Kley and A.~T\"unnermann for fruitful discussions.

\end{document}